\documentclass{article}
\usepackage{graphicx}

\usepackage{amsmath}
\usepackage{amssymb}

\title{A2DMN: Anatomy-Aware Dilated Multiscale Network for Breast Ultrasound Semantic Segmentation\thanks{Corresponding Author: Min Xian(mxian@uidaho.edu)}}

\author{Kyle Lucke$^{1}$ Aleksandar Vakanski$^{1,2}$ Min Xian$^{1}$ \\ 
\small $^{1}$ Department of Computer Science, University of Idaho, Idaho Falls, ID 83402, USA \\
\small $^{2}$ Department of Nuclear Engineering and Industrial Management,  \\ 
\small University of Idaho, Idaho Falls, ID 83402, USA
}

\date{}

\begin{document}

\maketitle

\begin{abstract}
In recent years, convolutional neural networks for semantic segmentation of breast ultrasound (BUS) images have shown great success; however, two major challenges still exist. 1) Most current approaches inherently lack the ability to utilize tissue anatomy, resulting in misclassified image regions. 2) They struggle to produce accurate boundaries due to the repeated down-sampling operations. To address these issues, we propose a novel breast anatomy-aware network for capturing fine image details and a new smoothness term that encodes breast anatomy. It incorporates context information across multiple spatial scales to generate more accurate semantic boundaries. Extensive experiments are conducted to compare the proposed method and eight state-of-the-art approaches using a BUS dataset with 325 images. The results demonstrate the proposed method significantly improves the segmentation of the muscle, mammary, and tumor classes and produces more accurate fine details of tissue boundaries.
\end{abstract}


\section{Introduction} 

Although most current work focuses on binary tumor segmentation of breast ultrasound (BUS) images, there is an increasing interest in segmenting and understanding the whole breast tissues~\cite{ref:huang-trustworthy}. First, the surrounding tissues of breast tumors may show signs of malignancy. Second, the spatial relationship between a breast tumor and other breast tissues contributes to the accurate detection of breast cancer. E.g., if a tumor invades the chest wall in the BUS image, the patient would have a high risk of cancer. Third, accurate segmentation of the mammary layer is necessary for estimating breast density. Despite the success of the semantic segmentation approaches~\cite{ref:segnet} on natural images, most methods struggle to produce accurate segmentation maps with fine details.

The main contributions of this work are as follows. 1) We propose a new smoothness loss to encode breast anatomy, which encourages close pixels with similar intensities to be classified into the same semantic class. This strategy results in smooth transitions between tissues in the semantic map. 2) We present a novel architecture for performing semantic segmentation of whole breast tissues in BUS images. The proposed architecture utilizes dilated convolutions~\cite{ref:yu-multi} to incorporate spatial context information at multiple scales, resulting in a model that produces more accurate semantic boundaries.

\begin{figure*}
    \centering
    \includegraphics[width=\linewidth]{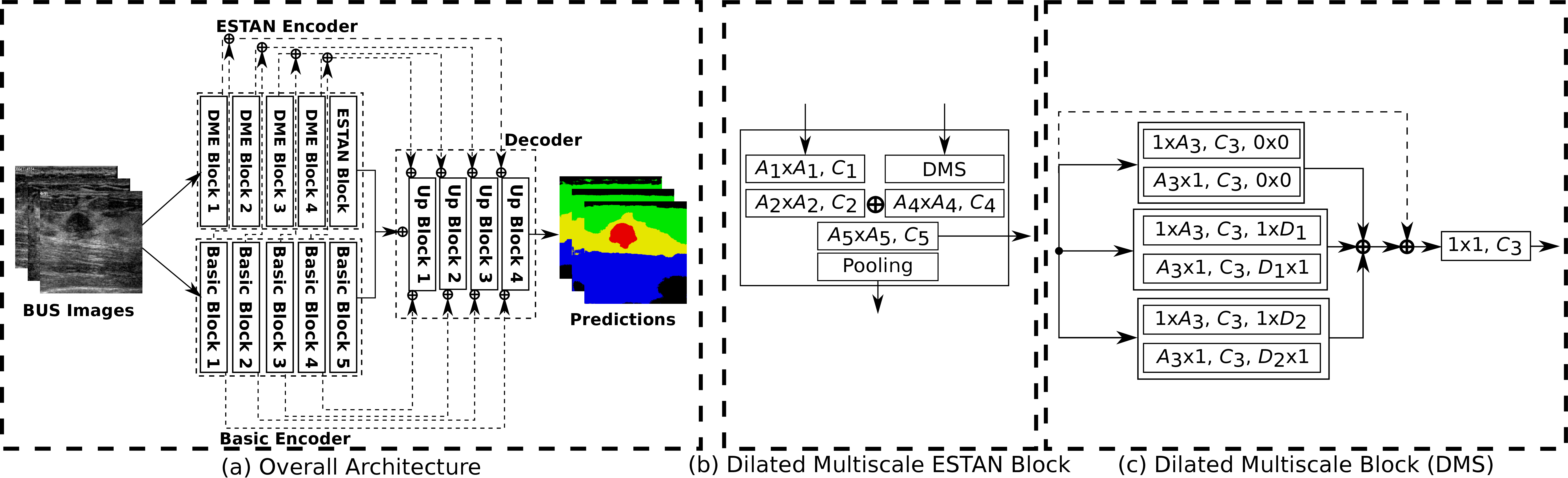}
    \caption{\textbf{A2DMN architecture.} \textbf{$\oplus$} is the concatenation operator. $A$ denotes kernel size, $C$ defines the number of kernels, $D$ denotes the dilation rate. DME blocks represent a Dilated Multiscale ESTAN block.}
    \label{fig:arch}
\end{figure*}

\section{Related Work} 

 Most conventional BUS segmentation approaches rely on classic image processing techniques~\cite{moon2014tumor}, graph-based techniques~\cite{ref:xian2015fully}, machine learning algorithms~\cite{pons2014computerized}, and level set methods~\cite{liu2010probability}. The approaches are typically binary segmentation and lack understanding of the whole breast tissues. Recently, deep learning-based segmentation approaches have dominated breast tumor segmentation in BUS images~\cite{ref:busis}; and most approaches are convolutional neural networks (CNNs)~\cite{ref:denseunet,ref:multiresunet,ref:rdau} and variants of U-Net~\cite{ref:u-net}.

 One popular technique for CNNs is to employ a fully connected conditional random field (CRF)~\cite{ref:crf} as a post-processing step to enhance smoothness. Due to their fully connected nature, CRFs can utilize context information and can efficiently resolve these ambiguities without adding any extra network parameters. However, CRFs cannot solve this issue in an end-to-end manner. CRF-RNN~\cite{ref:crf-rnn} solved this problem using a recurrent network (RNN) but added extra model parameters. Moreover, RNNs are not easy to parallelize on the GPU. It is well known that CNNs have trouble segmenting fine details and small objects because of the repeated down-sampling operations. Properly resolving these small objects and fine details is important when performing tumor segmentation since failure to adequately segment small tumors results in false negatives, which could be fatal. Dilated convolutions~\cite{ref:yu-multi} (DCs) utilize context information at multiple scales with little computational overhead. Most other works apply a single block comprised of several parallel DCs with increasingly large dilation rates to feature maps that have already been down-sampled~\cite{ref:rdau}. These feature maps may not contain enough relevant high-level information. 

\section{Methods}

\subsection{Dataset}\label{ssec:dataset}
The dataset used contains 325 BUS images collected from the First and Second Affiliated Hospitals of Harbin Medical University. Each image has been annotated by an experienced radiologist across five different classes representing subcutaneous fat, mammary, muscle, tumor, and background. The privacy of the patients is well protected. Data augmentation is utilized, and transformations include random rotations between $[-20, 20]$ degrees, horizontal flips, and random translations $\in [-5, 5]$ pixels. For pre-training, we use images from three BUS datasets labeled for binary tumor segmentation: BUSIS dataset~\cite{ref:busis} (562 images), UDIAT dataset~\cite{ref:b-dataset} (163 images), and BUSI dataset~\cite{ref:busi} (647 images). Note that the original BUSI dataset contains 133 images that do not contain tumors; we do not use these images for pre-training. To ensure the geometry of the anatomy is not distorted, all images are zero-padded to be square before resizing. Pixels added for padding are labeled as background in the segmentation mask.

\subsection{Network Architecture}
 In this work, we propose a new network architecture for the semantic segmentation of BUS images. It applies ESTAN~\cite{ref:estan} as the backbone network. ESTAN is an encoder-decoder segmentation network based on U-Net which has been specialized to perform binary tumor segmentation. We extend ESTAN to perform semantic segmentation and propose a new network architecture, shown in Figure~\ref{fig:arch}(a).

\textbf{Basic Block}. Each block, besides block 5, consists of two 3x3 convolutions, followed by a max pooling operation. The number of kernels for both convolutions in block 1, 2, 3, 4, and 5 are 32, 64, 128, 256, and 512, respectively.  
 
\textbf{Dilated Multiscale ESTAN (DME) Block}. The proposed DME is shown in Figure~\ref{fig:arch}(b). The DMS block (Figure~\ref{fig:arch}(c)) consists of three parallel DCs with different dilation rates and a residual connection. The resulting feature maps are then concatenated together, and a 1x1  convolution operation is used to project the concatenated feature maps back to the initial number of feature maps. By using DCs with different dilation rates, the network captures richer spatial and semantic context information at varying scales, thus allowing the network to accurately segment objects of varying shapes and sizes. For all blocks, $D_1$ is 2, and  $D_2$ is 4. The size of $A_3$ in DME blocks 1, 2, 3, and 4 are 15, 13, 11, and 9, respectively. The number of kernels ($C_i$) in blocks 1, 2, 3, and 4 is 32, 64, 128, and 256, respectively. The size of $A_5$ in blocks 2 and 5 is 5 and 1 in the rest of the blocks. 

\textbf{Up Block}. Each Up block consists of one up-sampling convolution layer followed by three convolutional layers. Let $f_{M_i, Y_i}$ denote a convolution with kernel size $M_i$ and $Y_i$ kernels. The output of the $j^{th}$ decoder block is given by $U_j = f_{M_3, Y_3}( f_{M_2, Y_2} ( f_{M_1, Y_1}(\Psi) ) )$ where $\Psi$ is the up-sampling layer. $M_1$, $M_3$ = 3 for all blocks, $M_2$ = 1 for blocks 1, 2, and 3, $M_2$ = 5 in block 4, and $Y_1$, $Y_2$, $Y_3$ = 256, 128, 64, and 32 in blocks 1, 2, 3, and 4 respectively. There are also two skip connections. The first skip connection combines the results of the first convolution in the basic block with the result of $f_{A_5, C_5}$. The second connection concatenates the results of the second convolution of the basic block with $f_{M_2, Y_2}$. The output layer utilizes a 1x1 convolution with five kernels, followed by a softmax activation. 

\subsection{Smoothness Loss}

Current loss functions lack control over the smoothness of the predicted segmentation maps, which results in many small misclassified image regions in the final semantic maps. We address the issue by proposing a new smoothness loss term to encode breast anatomy and encourage local pixel smoothness. The total loss used to train the network is $\mathcal{L}_{total} = \mathcal{L}_{\textrm{Dice}} + \lambda \mathcal{L}_{\textrm{SL}}$, where $\mathcal{L}_{\textrm{Dice}}$ is the Dice loss, $\mathcal{L}_{\textrm{SL}}$ is our proposed smoothness loss, and $\lambda$ balances the two loss terms. We use the Dice loss to mitigate problems of imbalanced classes in the training dataset. The smoothness loss is defined by 

\begin{equation*}
  \mathcal{L}_{\textrm{SL}} = \frac{1}{N} \sum_{i}^{\textrm{N}} \frac{\sum_{j \in \mathbb{N}_i} \Bigg[ \delta (V_i, V_j) \cdot \overbrace{ \exp\bigg( -\frac{ (I_{i} - I_{j})^2 }{\sigma_{\alpha}} \bigg) }^{\textrm{pixel smoothness}} \cdot \overbrace{\exp\bigg( \frac{ (V_{i} - V_{j})^2 }{\sigma_{\beta}} \bigg) }^{\textrm{label smoothness}} \frac{1}{d_{ij}}\Bigg]} {\sum_{j \in \mathbb{N}_i} [ \delta(V_i, V_j) ] + \epsilon} \textrm{,}
\end{equation*}

\noindent where $I_i$ and $I_j$ is the intensity value of the $i^{\textrm{th}}$ and $j^{\textrm{th}}$ image pixel, respectively; $V_i$ is the predicted class label for the $i^{\textrm{th}}$ pixel; $d_{ij}$ is the Euclidean distance between pixels $i$ and $j$; $N$ is the number of pixels in an image, $\mathbb{N}_i$ is the set of the eight immediate neighbors for pixel $i$; $\sigma_{\alpha}$ and $\sigma_{\beta}$ control the steepness of the curves; $ \delta (V_i, V_j)$ returns 1 if $V_i \neq V_j$ or 0 otherwise; and $\epsilon$ is a small positive number used to prevent division by zero. To obtain class labels that are still differentiable, we apply the softargmax function to the network outputs.

The pixel smoothness term encourages the network to classify nearby pixels with similar intensities into the same category by giving high penalties to similar pixels with different class labels. The label smoothness term encodes anatomical information for different types of neighboring tissues by giving high penalties to tissue layers that should not be found near one another.

\subsection{Transfer Knowledge From Binary Segmentation}\label{ssec:tl}
The dataset used in this work only has 325 images, making it difficult to train a deep network adequately. To address this challenge, BUS images with binary ground truths from three datasets are involved to pre-train all networks. For pre-training, we modify each of the networks to perform binary segmentation by replacing the final layer of each network with a single channel 1x1 convolution followed by a sigmoid activation function. Each network is optimized according to the Dice loss. Once the network for binary segmentation is trained, the encoder weights are used to initialize the encoder of each respective semantic segmentation network.

\section{Results}

 \subsection{Training Settings and Evaluation Metrics}

The networks are trained for 100 epochs using the Adam optimization method with a learning rate of $1e^{-4}$ and a batch size of 32. The weights for the best model according to the validation loss are used for testing. For all experiments which involve smoothness loss, $\sigma_{\alpha}$ = 0.1, $\sigma_{\beta}$ = 5, $\lambda = 5e^{-5}$, and $\epsilon = 1e^{-7}$. For the softargmax function, we set $\beta = 1e^{10}$. Each image is resized to 256x256 pixels and pixel values are scaled to $[0, 1]$. Due to the small size of the dataset, we validate the performance of our model using 5-fold cross-validation. All reported results in this section are averaged over all five folds. 15\% of the training data is used for validation. To measure the performance of our method, we employ one area metric and two boundary metrics: Intersection over Union (IoU), Hausdorff Distance (HD), and Average Absolute Distance (AAD). IoU measures the percentage of area overlap between the predicted and ground truth segmentation maps, while HD and AAD measure the maximum and average error between the predicted and ground truth segmentation boundaries in pixels, respectively.

\begin{table}
\centering
\begin{tabular}{|l|l|l|l|l|l|}
\hline
\textbf{Method}   & 
\textbf{BG} &
\textbf{MU}  & 
\textbf{F} & 
\textbf{MA} & 
\textbf{T} \\ \hline 
A2DMN w/ EB  & 88.7          & 71.9          & 81.7          & 75.2          & 80.5 \\ \hline
A2DMN w/ DB  & \textbf{90.6} & \textbf{74.9} & \textbf{83.2} & \textbf{76.7} & \textbf{80.6} \\ \hline
\end{tabular}
\caption{IoU values with and without DME blocks. BG: background, MU: muscle, F: fat, MA: mammary, T: tumor, EB: ESTAN Blocks, DB: DME Blocks. Best values in bold. Higher values are better.}\label{tab:dme-area}
\end{table}

\begin{table}
    \centering
\begin{tabular}{|l|l|l|l|l|l|}
\hline
\multicolumn{6}{|c|}{\textbf{HD (PX)}}\\ \hline
\textbf{Method} & \textbf{BG} & \textbf{MU} & \textbf{F}  & \textbf{MA} & \textbf{T} \\ \hline

A2DMN w/ EB & 33.8 & 34.8          & 27.7          & 26.9                   & \textbf{23.4} \\ \hline
A2DMN w/ DB   & \textbf{33.0} & \textbf{30.3} & \textbf{20.9} & \textbf{24.1} & 24.1 \\ \hline

\multicolumn{6}{|c|}{\textbf{AAD (PX)}}\\ \hline
A2DMN w/ EB & 2.8          & 7.00          & 3.5          & 4.4          & \textbf{8.2} \\ \hline
A2DMN w/ DB   & \textbf{2.6} & \textbf{6.00} & \textbf{2.8} & \textbf{3.8} & 8.4 \\ \hline
    \end{tabular}
    \caption{HD and AAD with and without DME blocks. PX: pixels. Lower values are better.}
    \label{tab:dme-boundary}
\end{table}

\subsection{The Effectiveness of the DME Blocks}
To verify the effectiveness of the DME block, we test the network with ESTAN blocks and with DME blocks in Tables~\ref{tab:dme-area} and \ref{tab:dme-boundary}. The experimental results in Table~\ref{tab:dme-area} show that replacing ESTAN blocks with our DME blocks results in a 1.9\%, 3.0\%, 1.5\%, and 1.5\% increase in IoU for the background, muscle, fat, and mammary classes, respectively, and remains competitive with respect to the tumor segmentation category. The experimental results in Table~\ref{tab:dme-boundary} demonstrate a reduction in HD by 4.5, 6.8, and 2.8 pixels in the muscle, fat, and mammary tissue classes, respectively, with competitive performance in all other categories. Meanwhile, the AAD is slightly reduced for all tissue categories, except the tumor category. 

\begin{table}
    \centering
\begin{tabular}{|l|l|l|l|l|l|}
\hline
\multicolumn{6}{|c|}{\textbf{HD (PX)}}\\ \hline
\textbf{Method} & \textbf{BG} & \textbf{MU} & \textbf{F}  & \textbf{MA} & \textbf{T} \\ \hline
A2DMN w/o SL & 33.0          & 30.4          & 20.9          & 24.1          & 24.1          \\ \hline
A2DMN w/  SL & \textbf{30.2} & \textbf{25.7} & \textbf{19.1} & \textbf{21.4} & \textbf{20.1} \\ \hline
\multicolumn{6}{|c|}{\textbf{AAD (PX)}}\\ \hline
A2DMN w/o SL & 2.6          & 6.0          & 2.8          & 3.8          & 8.4 \\ \hline
A2DMN w/  SL & \textbf{2.4} & \textbf{5.8} & \textbf{2.7} & \textbf{3.7} & \textbf{6.9} \\ \hline
    \end{tabular}
    \caption{Comparison of the proposed method with and without the smoothness loss, using the boundary metrics HD and AAD. Lower values are better.}
    \label{tab:sl-boundary}
\end{table}

\begin{table}
    \centering
\begin{tabular}{|l|l|l|l|l|l|}
\hline
\multicolumn{6}{|c|}{\textbf{HD (PX)}}\\ \hline
\textbf{Method} & \textbf{BG} & \textbf{MU} & \textbf{F}  & \textbf{MA} & \textbf{T} \\ \hline
RDAU-NET        & 61.1          & 97.3          & 59.2          & 52.2          & 95.2          \\ \hline
CE-Net          & 31.1          & 33.1          & 26.3          & 27.0          & 51.4          \\ \hline
DenseU-Net      & 69.0         & 106.6         & 77.6          & 77.6          & 120.8         \\ \hline
MultiResUNet    & 56.0          & 93.1          & 99.7          & 73.5          & 115.6         \\ \hline
SCAN            & 34.2          & 30.0          & 21.8          & 26.3          & 30.8          \\ \hline
SegNet          & 63.0          & 78.0          & 75.7           & 59.0          & 91.9          \\ \hline
ESTAN           & 33.8          & 34.8          & 27.7          & 26.9          & 23.4          \\ \hline
SCFURNet        & \textbf{30.0} & 28.6          & 25.8          & 22.9          & 21.9          \\ \hline
A2DMN           & 30.2          & \textbf{25.7} & \textbf{19.1} & \textbf{21.4} & \textbf{20.1} \\ \hline

\multicolumn{6}{|c|}{\textbf{AAD (PX)}}\\ \hline
RDAU-NET     & 8.3          & 22.0          & 12.1          & 14.9          & 36.7         \\ \hline
CE-Net       & 3.3          & 8.2           & 3.9           & 6.2           & 20.2         \\ \hline
DenseU-Net   & 10.5         & 29.8          & 17.2          & 16.9          & 50.2         \\ \hline
MultiResUNet & 8.1          & 24.5          & 28.9          & 19.9          & 47.2         \\ \hline
SCAN         & 2.9          & 6.3           & 3.0           & 4.3           & 10.0         \\ \hline
SegNet       & 8.3          & 16.4          & 11.2          & 13.6          & 40.7         \\ \hline
ESTAN        & 2.8          & 7.0           & 3.5           & 4.4           & 8.2          \\ \hline
SCFURNet     & 2.8          & 6.4           & 3.6           & 4.2           & 7.5          \\ \hline
A2DMN        & \textbf{2.4} &  \textbf{5.8} & \textbf{2.7}  & \textbf{3.7}  & \textbf{6.9} \\ \hline
    \end{tabular}
    \caption{Comparison of the proposed method and other methods using the boundary metrics HD, and AAD.}
    \label{tab:sota-boundary-metrics}
\end{table}

\subsection{The Effectiveness of the Smoothness Loss}
\begin{figure}
\centering
\includegraphics[width=\linewidth]{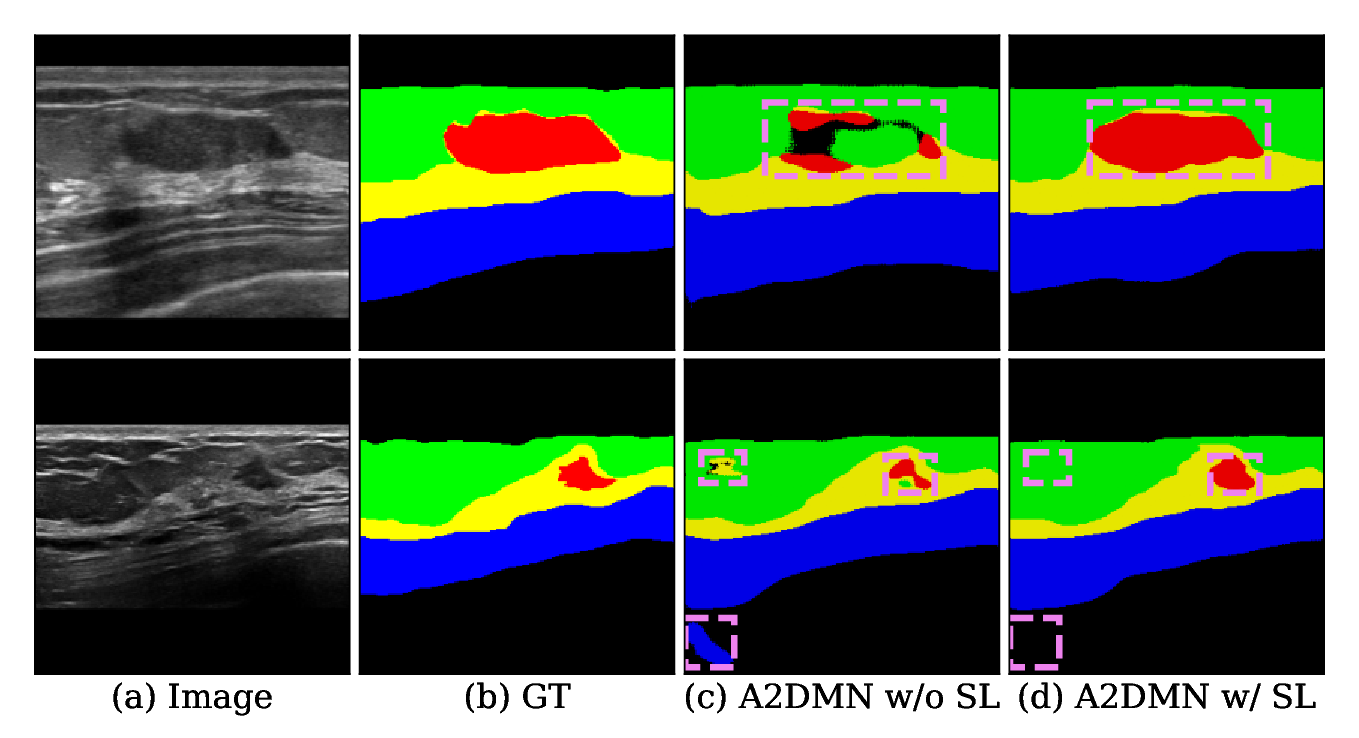}
\caption{A2DMN with and without the proposed smoothness loss. The violet dashed boxes represent misclassified regions.}
\label{fig:sl-qual}
\end{figure}

To verify the performance of the smoothness loss, we test the proposed network with and without $\mathcal{L}_{SL}$ in Table~\ref{tab:sl-boundary}. The addition of the smoothness loss results in a reduction of HD by 2.8, 4.7, 2.7, and 4.0 pixels in the background, muscle, mammary, and tumor categories, respectively. Table~\ref{tab:sl-boundary} also shows an improvement of 1.5 pixels in the tumor category and a slight reduction in AAD in all other tissue categories. The results demonstrate that the smoothness loss improves performance and generates more accurate results for the most critical tissue. As shown in Figure~\ref{fig:sl-qual}, the most common misclassifications (dashed boxes) are heterogeneous image regions correctly classified when the smoothness loss is applied.

\subsection{Overall performance}\label{sec:overall}

Eight state-of-the-art methods, ESTAN, CE-Net\cite{ref:cenet}, DenseU-Net\cite{ref:denseunet},  \\ MultiResUNet\cite{ref:multiresunet}, SCAN\cite{ref:scan}, SegNet\cite{ref:segnet}, RDAU-NET\cite{ref:rdau}, and SCFURNet\cite{ref:huang-trustworthy} are compared with the proposed method. The results are shown in 
Table \ref{tab:sota-boundary-metrics}. Under the HD metric, we 
outperform every other model by at least 2.9, 2.7, 1.5, and 1.8 pixels in the muscle, fat, mammary, and tumor 
categories, respectively, and are competitive in every other category. Under the AAD metric, we are competitive in all 
categories, slightly outperforming the next best-performing models in every class but background.

\section{Conclusion}
In this paper, we propose a novel approach for the semantic segmentation of breast ultrasound images, namely A2DMN, which consists of two major components: a dilated multiscale network architecture and a smoothness loss. The proposed method achieves state-of-the-art performance for muscle, mammary, and tumor segmentation, and competitive performance in segmenting the background and fat categories; and it produces finer details around tissue boundaries. Good performance is achieved for two reasons. 1) The proposed smoothness loss term encodes breast anatomy and pixel smoothness, resulting in smoother, more anatomically accurate segmentation maps with little computational overhead. 2) The proposed network uses the DME blocks and can provide better context information at multiple scales, allowing the network to segment semantic boundaries accurately.

\section{Compliance with Ethical Standards}
This research study was conducted retrospectively using human subject data made available in open access (http://cvprip.cs.usu.edu/busbench/). Ethical approval was not required, as confirmed by the license attached with the open-access data.

\section{Acknowledgments}
Research reported in this publication was partially supported by the National Institute Of General Medical Sciences of the National Institutes of Health under Award Number P20GM104420.

\bibliographystyle{unsrt}
\bibliography{refs}

\end{document}